\documentclass[aps,prb,superscriptaddress,twocolumn,showkeys,amsmath,amssymb,floatfix]{revtex4-2}
\usepackage[colorlinks=true,linkcolor=Blue,allcolors=blue]{hyperref}
\usepackage{graphicx}% Include figure files
\usepackage{dcolumn}% Align table columns on decimal point
\usepackage{bm}% bold math
\usepackage{natbib}
\usepackage{soul}

\begin{document}

\preprint{APS/123-QED}

\title{Spin-split collinear antiferromagnets: a large-scale ab-initio study}
\author{Yaqian Guo}
 \affiliation{Leibniz Institute for Solid State and Materials Research, IFW Dresden, Helmholtzstraße 20, 01069 Dresden, Germany}
\author{Hui Liu}
 \affiliation{Leibniz Institute for Solid State and Materials Research, IFW Dresden, Helmholtzstraße 20, 01069 Dresden, Germany}
  \affiliation{W\"urzburg-Dresden Cluster of Excellence ct.qmat, Technische Universit\"at Dresden, 01062 Dresden, Germany}
 \author{Oleg Janson}
 \affiliation{Leibniz Institute for Solid State and Materials Research, IFW Dresden, Helmholtzstraße 20, 01069 Dresden, Germany}
 \author{Ion Cosma Fulga}
 \affiliation{Leibniz Institute for Solid State and Materials Research, IFW Dresden, Helmholtzstraße 20, 01069 Dresden, Germany}
  \affiliation{W\"urzburg-Dresden Cluster of Excellence ct.qmat, Technische Universit\"at Dresden, 01062 Dresden, Germany}
 \author{Jeroen van den Brink}
 \affiliation{Leibniz Institute for Solid State and Materials Research, IFW Dresden, Helmholtzstraße 20, 01069 Dresden, Germany}
 \affiliation{W\"urzburg-Dresden Cluster of Excellence ct.qmat, Technische Universit\"at Dresden, 01062 Dresden, Germany}
 \author{Jorge I. Facio}
 \email{jorge.facio@cab.cnea.gov.ar}
 \affiliation{Leibniz Institute for Solid State and Materials Research, IFW Dresden, Helmholtzstraße 20, 01069 Dresden, Germany}
  \affiliation{Centro At\'omico Bariloche and Instituto Balseiro, CNEA, 8400 Bariloche, Argentina}
  \affiliation{Instituto de Nanociencia y Nanotecnolog\'ia CNEA-CONICET, 8400 Bariloche Argentina}

\begin{abstract}
It was recently discovered that, depending on their symmetries, collinear antiferromagnets can actually break the spin degeneracy in momentum space, even in the absence of spin-orbit coupling. 
Such systems, recently dubbed \textit{altermagnets}, are signalled by the emergence of a spin-momentum texture set mainly by the crystal and magnetic structure, relativistic effects playing a secondary role. 
Here we consider all collinear $q$=0 antiferromagnetic compounds in the MAGNDATA database allowing for spin-split bands. 
Based on density-functional calculations for the experimentally reported crystal and magnetic structures, we study more than sixty compounds and introduce numerical measures for the average momentum-space spin splitting. We highlight some compounds that are of particular interest, either due to a relatively large spin splitting, such as CoF$_2$ and FeSO$_4$F, or because of their low-energy electronic structure. The latter include LiFe$_2$F$_6$, which hosts nearly flat spin-split bands next to the Fermi energy, as well as RuO$_2$, CrNb$_4$S$_8$, and CrSb, which are spin-split antiferromagnetic metals.
\end{abstract}

%Use showkeys class option if keyword
\maketitle

%\tableofcontents
\section{INTRODUCTION}
A basic characteristic of translationally-invariant electronic systems is the single-electron energy dispersion, $\varepsilon_{n,\mathbf{k},\sigma}$, where $n$ is a band index, $\mathbf{k}$ the crystal momentum and $\sigma$ the electron spin. Two well-known mechanisms for breaking the spin degeneracy ($\varepsilon_{n,\mathbf{k},\uparrow}\neq\varepsilon_{n,\mathbf{k},\downarrow}$) are the Zeeman effect due to a nonzero net magnetization and the Rashba-Dresselhaus effect caused by the spin-orbit coupling (SOC) in noncentrosymmetric materials. Recent works have studied the general symmetry conditions that allow for a broken spin degeneracy \cite{Yuan, Yuan2020Giant, egorov2021antiferromagnetism}, finding that mechanisms fundamentally different from the Zeeman effect and SOC can still lead to spin-split bands. 
In particular, under certain symmetry conditions, collinear antiferromagnets (cAFMs) can break the spin degeneracy in momentum space even in the absence of SOC. The energy scale of this spin-splitting is set by the local exchange field, which is orders of magnitude larger than typical field-induced Zeeman splittings, reaching maximum values that can substantially exceed the SOC energy scale. Such antiferromagnets naturally dispense with heavy elements and avoid problems associated with stray fields, making them promising candidates for various spintronics applications \cite{Rafael2021PRL, Naka2021PRB,shao2021spin, Libor2022PRX}. These ideas have contributed to the recent resurgent interest in various aspects of cAFMs \cite{vsmejkal2020crystal, hayami2019momentum, Hayami2020PRB, mazin2021prediction, bose2021tilted, bai2021observation, egorov2021colossal, smejkal2021anomalous}. 

Magnetic space groups (MSGs) are usually classified into four types, depending on how they are constructed in relation to the parent crystallographic space group $G$ \cite{bradley2010mathematical}. Type I MSGs, $M_{\rm I}$, are equivalent to their parent group $G$, which only contains unitary symmetries, hence $M_{\rm I}$ = $G$. Such systems break not only time-reversal symmetry ($\theta$), but also any combination of a spatial symmetry with $\theta$. Particularly relevant is the combination of $\theta$ and inversion symmetry ($I$). Under these operations the energy dispersion transforms as
\begin{equation}
\begin{split}
\theta : \varepsilon_{n,\mathbf{k},\sigma} & \to \varepsilon_{n,-\mathbf{k},-\sigma}, \\
I : \varepsilon_{n,\mathbf{k},\sigma} & \to \varepsilon_{n,-\mathbf{k},\sigma}.
\end{split}
\end{equation}
Thus, breaking their product, $\theta I$, is necessary to obtain a spin splitting in momentum space, and all materials of type-I MSG allow for spin-split bands in the limit of vanishing SOC. Type II MSGs, $M_{\rm II}$, contain all the symmetry operations of $G$ plus their combination with $\theta$, i.e. $M_{\rm II}$ = $G$ + $\theta G$. This corresponds to nonmagnetic phases, which we do not consider in this work. MSGs of type III, $M_{\rm III}$, preserve a halving subgroup of $G$ called $H$ while the remaining operations are replaced by their combination with $\theta$, namely $M_{\rm III}$ = $H$ + $\theta (G-H)$, where ($G-H$) does not contain pure translations. MSGs of type IV, $M_{\rm IV}$, consist of all the elements of G plus an equal number of antiunitary operations, each of which is an operation of $G$ multiplied by $\theta$ and a translation operation $T$, i.e. $M_{\rm IV}$ = $G$ + $\theta TG$. $T$ connects the two sublattices of opposite magnetic moment, and therefore $T\theta$ is a symmetry element in $M_{\rm IV}$, also called antitranslation symmetry. The modified translations reflect the fact that the magnetic primitive cell in type IV compounds becomes a supercell of the parent primitive cell. Consequently, ions with opposite spin have equivalent local environments and the system is symmetric under the combination of $T$ with a pure spin rotation, $U_s$, which maps the spin $\uparrow$ to $\downarrow$ and vice versa. For a vanishing SOC, the spin-degeneracy of all bands in type IV MSGs is protected by $U_s T$.
In  summary, a finite spin splitting in the absence of SOC can only occur in compounds having a type I or III MSG \cite{Yuan, Yuan2020Giant}, and several examples have already been analyzed in Refs.~\cite{Yuan, Yuan2020Giant}. 

When analyzed in terms of non-relativistic symmetry groups, spin-split cAFMs can be formally classified as distinct from ferromagnets and from antiferromagnets having spin-degenerated bands \cite{jungwirth2021altermagnetism}. In this classification, they realize a phenomenon recently dubbed \textit{altermagnetism}, whose  hallmark are non-trivial spin-momentum textures yielding in total a net zero magnetization \cite{vsmejkal2022emerging}.

In this work, we consider all compounds whose magnetic structure has been experimentally studied and included up to date in the MAGNDATA database, a collection of more than a thousand published magnetic structures \cite{gallego2016magndata, gallego2016magndata2, perez2015symmetry}. We identify those cAFMs that fulfill the symmetry conditions required for spin-split bands in the absence of SOC. We filter out particularly complicated systems (e.g. systems where the long-range magnetic order stems from rare earth atoms or compounds with many atoms in the unit cell) and perform density-functional calculations assuming the experimental crystal and magnetic structures. We introduce measures to systematically estimate and compare the average spin splitting in momentum space. Our calculations allow us to identify those compounds which may be particularly suited for experimentally studying the physics associated with spin-split electronic bands. In addition, we discuss in some detail a number of compounds which are especially interesting, either because of the magnitude of the spin splitting, or because of their low-energy (near Fermi energy) electronic structure. Last, taking as example the case of RuO$_2$, we discuss the possible effects of the SOC on the band structure of spin-split AFMs.

\section{AB-INITIO METHODOLOGY}
The results reported in this work are based on density-functional calculations (DFT) performed in the scalar-relativistic approximation. 
We use the generalized gradient approximation (GGA) for the exchange and correlation functional and treat the partially filled atomic shells of transition metal atoms with the $+U$ method,  using the fully localized limit functional for the double-counting correction, as implemented in the FPLO code \cite{Kopernik1999}. 
The values of the Hubbard $U$ parameter are taken from previous studies and are indicated in Tables \ref{compounds} and  \ref{compounds2}.
A linear $k$-mesh density of 50\AA$^{-1}$ is employed for the Brillouin zone (BZ) integrations, using the tetrahedron method. 
For instance, if the real space lattice constants are $a=b=c=5$\AA, a $k$-mesh having $10\times10\times10$ subdivisions is used. 
Lastly, we use as convergence criteria that the charge density is converged below $10^{-6}$ $a_B^{-3}$ (with $a_B$ the Bohr radius) or that the total energy is converged below $10^{-6}$ Hartree.

Our calculation scheme consists of three steps, where the charge density from the previous step is used to facilitate the convergence of the subsequent calculation. 
The order of the calculations is as follows: (1$^{\rm st}$) nonmagnetic scalar-relativistic, (2$^{\rm nd}$) magnetic scalar-relativistic, and (3$^{\rm rd}$)  magnetic scalar-relativistic GGA$+U$ calculations. For some compounds, we also performed fully-relativistic calculations using the four-component formalism.
Lastly, for a quantitative analysis of the momentum-space band splitting, we constructed Wannier functions-based tight-binding Hamiltonians using the automatic symmetry-conserving projective method implemented in FPLO \cite{koepernik2021symmetry}. The band structure of the Wannier model and that of the DFT calculation tipically differ in less than 10\,meV. 
We use these models for the evaluation of Eqs. [\ref{eq_m1}-\ref{eq_m3}]. For these calculations, we applied a unified approach and chose the $k$-meshes such that they facilitate a possibly uniform sampling of the Brillouin zone and the number of points is close to the target value of $2\times10^4$ per atom. 
Convergence with respect to the latter value was carefully checked.
\begin{figure*}[htp!]
	\begin{center}
		\includegraphics[width=\textwidth]{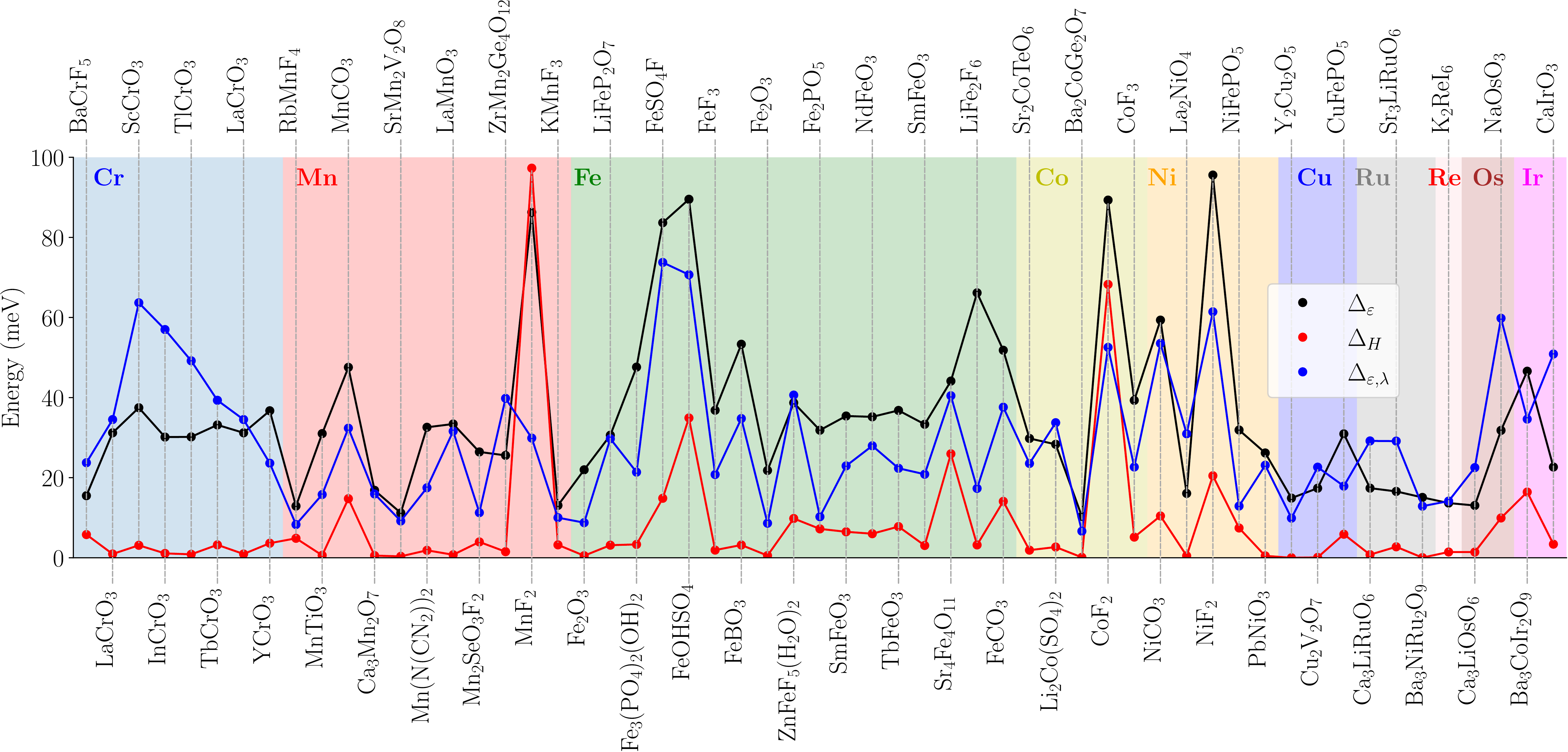}
		\caption{Measures of the average energy splitting  between opposite spin bands for all the compounds considered that result in insulators. Sectors of different color correspond to different magnetic ions. $\Delta_\varepsilon$, $\Delta_{H}$ and $\Delta_{\varepsilon, \lambda}$ are defined in Eqs.  [Eq.~\eqref{eq_m1}], 
		 [Eq.~\eqref{eq_m2}] and   [Eq.~\eqref{eq_m3}], respectively.}
		\label{fig_measures}
	\end{center}
\end{figure*}

\section{Compounds selection}
\label{sec_selection}
We use experimental magnetic and crystal structures collected from the MAGNDATA database \cite{magndata}. We scanned the MAGNDATA database searching for cAFMs that allow for spin splitting in the limit of zero SOC.
As explained above, these are either type I MSGs or those type III MSGs which additionally break $\theta I$. We restrict ourselves to compounds with a magnetic structure with a single null propagation vector (\textit{Class 0} according the MAGNDATA classification \cite{gallego2016magndata, gallego2016magndata2, perez2015symmetry}). These are compounds for which the translational symmetry of the magnetic structure and of the crystal structure are identical. At the time of performing the material selection, this class had 779 entries. Of this total, close to 38\% of the reported magnetic structures are noncollinear, approximately 24\%  have $\theta I$ as a symmetry and nearly 18\% have a ferromagnetic component (either explicitly included in the reported magnetic structure or indicated as an additional comment). 

In addition, about 5\% are either nonstoichiometric, or compounds in which the long-range magnetic order is associated with  magnetic rare-earth ions. For simplicity, we have excluded these types of compounds from the present work. 

We have additionally left out of the present study certain compounds whose DFT calculation has proven hard to converge, thus requiring further detailed research. These compounds are La$_2$LiRuO$_6$ \cite{Battle2003JSSC}, Fe$_2$Mo$_3$O$_8$ \cite{Mn2Mo3O8}, Co$_2$Mo$_3$O$_8$ \cite{Co2Mo3O8}, CaFe$_5$O$_7$ \cite{CaFe5O7}, Fe$_2$O$_3$ \cite{e-Fe2O3}, CsCoF$_4$ \cite{CsCoF4}, Sr$_3$NaRuO$_6$ \cite{Sr3LiRuO6}, Sr$_3$NiIrO$_6$ \cite{Sr3NiIrO6}, CsCoCl$_3$ \cite{CsCoCl3}, Ba$_5$Co$_5$ClO$_{13}$ \cite{Ba5Co5ClO13}, Sr$_2$LuRuO$_6$ \cite{Sr2LuRuO6}, KMnF$_3$ \cite{KMnF3} and YBaMn$_2$O$_5$ \cite{YBaMn2O5}. 

In addition, we found that some compounds whose DFT calculations show poor convergence, tend to break some of the antiunitary symmetries present in the experimentally reported crystal and magnetic structure, symmetries which are not enforced during the computation. In such cases, even if the net magnetic moment, i.e.\ the difference between the spin-up and spin-down densities integrated up to the Fermi energy, is approximately zero, the densities of states of the two spin channels have a different energy dependence. 
As a result, a shift of the Fermi energy readily gives rise to a net magnetization. While such instability towards breaking the antiferromagnetic order could be of physical interest, we have excluded these compounds from the present work. These excluded compounds are MnLaMnSbO$_6$ \cite{MnLaMnSbO6}, CoCO$_3$ \cite{CoCO3}, Li$_3$Fe$_2$(PO$_4$)$_2$ (MSG $P2_1^\prime/c^\prime$) \cite{Li3Fe2(PO4)3}, Li$_3$Fe$_2$(PO$_4$)$_2$ (MSG $R\bar{3}$) \cite{Li3Fe2(PO4)3}, Mn$_2$Mo$_3$O$_8$ \cite{Mn2Mo3O8}, BaMn$_2$Si$_2$O$_7$ \cite{BaMn2Si2O7}, Cr$_2$S$_3$ \cite{Cr2S3}, BiCrO$_3$ (MSG $P\bar{1}$) \cite{BiCrO3} and BiCrO$_3$ (MSG $C2/c$) \cite{BiCrO3}.

The remaining is a set of close to sixty compounds, presented in Tables \ref{compounds} and \ref{compounds2}. For these compounds, we have verified that the eigenvalues at $\Gamma$ appear in pairs of approximately equal energy.
Specifically, we have computed the average difference $\frac{1}{N_b}\sum_{n}|\varepsilon_{\Gamma,n,\uparrow}-\varepsilon_{\Gamma,n,\downarrow}|$, with $N_b$ the total number of bands, finding it  to be generally smaller than 6\,meV. As a side remark, our materials' list includes some compounds which have magnetic sublattices associated with two different transition-metal atoms. These are CuFePO$_5$, NiFePO$_5$, Ba$_3$NiRu$_2$O$_9$, Ba$_3$CoIr$_2$O$_9$ and Sr$_2$CoOsO$_6$.

Lastly, for over 90\% of the 57 compounds for which we found experimental data for the magnetic moment, the calculated magnetic moment agrees with the experimental values within 30\%. We conclude from this that the choice of the values $U$ are fairly reasonable. Notice that different factors, such as temperature, hybridization with ligands and quantum fluctuations, tend to reduce the magnetic moment. In  DFT$+U$, hybridization with ligands is not properly described, quantum fluctuations are severely underestimated and our calculations are at zero temperature. These effects contribute to the calculated values being typically larger than the experimental ones. Compounds for which the difference is larger than 30\% include cases where the temperature associated with the experimental estimation is rather large (so that the ordered moment is likely not be fully saturated), like  SmFeO$_3$ and FeF$_3$, and cases where the low dimensional structure may enhance quantum fluctuations, like RbMnF$_4$.

% % 
\begin{table*}
\footnotesize
	\caption{\label{compounds}Collinear AFM  allowing for spin-split bands obtained from the MAGNDATA database. The columns denote their corresponding crystal structures, magnetic point groups (MPGs), collected $U$ values from previous studies, calculated and measured magnetic moments, and the reference reporting the crystal and magnetic structures used in this work. CHE stands for the crystal Hall effect and RT for room temperature. Metallic compounds are accompanied by the symbol $^*$ next to their name.}
	\begin{ruledtabular}
	\scalebox{0.98}{
        \begin{tabular}{ccccccccc}
 %   \toprule %\midrule
    \hline
	S. No & Compound & Structure & MPG &MSG& U value  & CHE   & Local moment & Local moment    \\
	%\midrule
	&&&[Reference]&type&(eV)&\footnotesize{symmetry-}&\footnotesize{calculation}&\footnotesize{experiment} \\
	&&&&&& \footnotesize{allowed}&($\mu_B$)&($\mu_B$)\\

	\hline\centering

	 1 (0.303) & BaCrF$_5$ & $P2_1 2_1 2_1$ (19) & $2^\prime2^\prime2$ \cite{BaCrF5}& III & 3.4 \cite{CoF3u} & YES & 3.08 & 3.87 (60 K) \\ %\midrule

	\hline\centering

	 2 (0.323) & LaCrO$_3$ & $Pnma$ (62) & $mmm$ \cite{LaCrO3c} & I &3.5 \cite{LaCrO3} & NO& 3.01 & 2.51 (5 K)   \\ %\midrule

	\hline\centering

	 3 (0.307) & ScCrO$_3$ & $Pnma$ (62) & $mmm$ \cite{ACrO3} & I &2.0 \cite{ACrO3} & NO & 2.93 & 2.70 (1.5 K)   \\ %\midrule

	\hline\centering

	 4 (0.308) & InCrO$_3$ & $Pnma$ (62) & $mmm$ \cite{ACrO3}& I &2.0 \cite{ACrO3} & NO & 2.98 & 2.50 (1.5 K)   \\ %\midrule

	\hline\centering

	 5 (0.309) & TlCrO$_3$ & $Pnma$ (62) & $mmm$ \cite{ACrO3}& I &2.0 \cite{ACrO3} & NO & 2.97 & 2.46 (1.5 K)   \\ %\midrule

	\hline\centering

      6 (0.354) & TbCrO$_3$ & $Pbnm$ (62) & $m^\prime m^\prime m$ \cite{TbCrO3s}& III &3.0 \cite{TbCrO3u} & YES & 3.00 & 2.85 (4.2 K)   \\ %\midrule

	\hline\centering

      7 (0.417) & LaCrO$_3$ & $Pnma$ (62) & $m^\prime m^\prime m$ \cite{LACrO3s}& III &3.5 \cite{LaCrO3} & YES & 3.03 & 3.0 (0 K)   \\ %\midrule

	\hline\centering

	8 (0.586) & YCrO$_3$& $Pbnm$ (62) & $m^\prime m^\prime m$ \cite{YCrO3s}& III &4.0 \cite{YCrO3u} & YES& 3.04 & 2.96 (4.2 K)   \\ %\midrule

	\hline\centering

	9 (0.708) & CrNb$_4$S$_8^*$ & $P6_3/mmc$ (194) & $6^\prime/m ^\prime mm^\prime $ \cite{CrNb4S8s}& III & 6.0 \cite{CrNb4S8u} & NO& 3.46 & 1.5 (4.2 K)  \\ %\midrule

	\hline\centering

	10 (0.528) & CrSb$^*$ & $P6_3/mmc$ (194) & $6^\prime/m ^\prime mm^\prime $ \cite{CrSbs}& III & 0 \cite{Park2020} & NO& 3.22 & 2.5 (0 K)  \\ %\midrule
	
	\hline\centering
	
	11 (0.329) & RbMnF$_4$ & $P2_1/a$ (14) & $\bar{1}$ \cite{RbMnF4s}& I &3.9 \cite{RbMnF4u} & YES & 4.15 & 2.97 (1.5 K)   \\ %\midrule

	\hline\centering

	 12 (0.50) & MnTiO$_3$ & $R3c$ (161) & $m^\prime$ \cite{MnTiO3s}& III &3.9 \cite{MnTiO3u} & YES & 4.97 &3.9 (2 K)   \\ %\midrule

	\hline\centering

	13 (0.115) & MnCO$_3$ & $R\bar{3}c$ (167) & $2/m$ \cite{MnCO3s}& I &4.5 \cite{MnNCN2u} & YES & 5.06 &   \\ %\midrule

	\hline\centering

	14 (0.23) & Ca$_3$Mn$_2$O$_7$ & $Cmc2_1$ (36) & $m^\prime m2 ^\prime$ \cite{Ca3Mn2O7s}& III &4.0 \cite{Ca3Mn2O7u} & YES & 2.95 & 2.67 (10 K)   \\ %\midrule

	\hline\centering
	
	15 (0.62) & SrMn$_2$V$_2$O$_8$ & $I4_1cd$ (110) & $m^\prime m2 ^\prime$ \cite{SrMn2V2O8s}& III &4.0 \cite{SrMn2V2O8u} & YES & 5.06 & 3.99 (1.5 K)   \\ %\midrule

	\hline\centering	
	
	16 (0.131) & Mn(N(CN)$_2$)$_2$ & $Pnnm$ (58) & $m^\prime m^\prime m$ \cite{MnNCN22s}& III &4.5 \cite{MnNCN2u} & YES & 5.11 & 5.01 (1.7 K)   \\ %\midrule

	\hline\centering

	17 (0.1) & LaMnO$_3$ & $Pnma$ (62) & $m^\prime m^\prime m$ \cite{LaMnO3s}& III &2.0 \cite{LaMnO3u} & YES & 3.96 & 3.87 (1.4 K)  \\ %\midrule

	\hline\centering

	18 (0.755) & Mn$_2$SeO$_3$F$_2$ & $Pnma$ (62) & $m^\prime m^\prime m$ \cite{Mn2SeO3F2s}& III &4.6 \cite{Mn2SeO3F2s}&YES & 5.15 & 4.34 (1.5 K)  \\ %\midrule

	\hline\centering

	19 (0.229) & Ba$_2$MnSi$_2$O$_7$ & $P\bar{4}2_1m$ (113) & $\bar{4}2m$ \cite{Ba2MnSi2O7s}& I &6.0 \cite{Ba2MnSi2O7u} & NO & 5.14 & 4.1 (1.6 K)   \\ %\midrule

	\hline\centering

	 20 (0.315) & ZrMn$_2$Ge$_4$O$_{12}$ & $P4/nbm$ (125) & $4^\prime/mm^\prime m$ \cite{ZrMnGe4O12s}& III &5.0 \cite{Yuan2020Giant} & NO & 5.17 & 4.68 (1.6 K)   \\ %\midrule
	
	\hline\centering

	 21 (0.15) & MnF$_2$ & $P4_2/mnm$ (136) & $4^\prime/mm^\prime m$ \cite{MnF2s}& III &5.0 \cite{Yuan2020Giant} & NO& 5.25 & 4.6 (10 K)  \\ %\midrule

	\hline\centering
	
	22 (0.433) & KMnF$_3$ & $I4/mcm$ (140) & $4/mmm$ \cite{KMnF3}& I &4.58 \cite{KMnF3u} &NO &5.35 & 4.8 (10 K)  \\ %\midrule

	\hline\centering

	23 (0.66) & Fe$_2$O$_3$ & $R\bar{3}c$ (167) & $\bar{1}$ \cite{aFe2O3s}& I &3.8 \cite{Fe2O3u} & YES  &4.38 & 4.222 (10 K)   \\ %\midrule

	\hline\centering
	
	24 (0.83) & LiFeP$_2$O$_7$ & $P2_1$ (4) & $2$ \cite{LiFeP2O7s}& I &4.9 \cite{CoF3u} & YES & 4.71 & 4.60 (1.5 K)   \\ %\midrule

	\hline\centering	

	 25 (0.392) & Fe$_3$(PO$_4$)$_2$(OH)$_2$ & $P2_1/c$ (14) & $2/m$ \cite{Fe3PO42OH2s}& I &4.9 \cite{CoF3u} & YES & 4.04, 4.61 & 4.3 (1.5 K)  \\ %\midrule

	\hline\centering

	26 (0.128) & FeSO$_4$F & $C2/c$ (15) & $2 ^\prime/m^\prime$ \cite{FeSO4Fs}& III &4.3 \cite{FeSO4Fu} & YES & 4.65 & 4.32 (2 K)   \\ %\midrule

	\hline\centering

	27 (0.760) & FeOHSO$_4$ & $C2/c$ (15) & $2 ^\prime/m^\prime$ \cite{FeOHSO4s}& III&4.3 \cite{FeSO4Fu} & YES& 4.59 & 4.09  \\ %\midrule

	\hline\centering

	28 (0.335) & FeF$_3$ & $R\bar{3}c$ (167) & $2 ^\prime/m^\prime$ \cite{lee2018weak}& III&4.9 \cite{CoF3u} & YES & 5.02 &2.60 (RT)  \\ %\midrule

	\hline\centering

	29 (0.112) & FeBO$_3$ & $R\bar{3}c$ (167) & $2 ^\prime/m^\prime$ \cite{FeBO3s}& III&6.0 \cite{FeBO3u} & YES & 4.66 & 4.70
	 (77 K)   \\ %\midrule

	\hline\centering
	
	 30 (0.65) & Fe$_2$O$_3$ & $R\bar{3}c$ (167) & $2 ^\prime/m^\prime$ \cite{aFe2O3s}& III&3.8 \cite{Fe2O3u} & YES & 4.38 & 4.118 (300 K)   \\ %\midrule
	
		\hline\centering
	
	 31 (0.575) & ZnFeF$_5$(H$_2$O)$_2$ & $Imm2$ (44) & $mm2$ \cite{ZnFeF5H2O2}& I &4.9 \cite{CoF3u} & NO& 4.76 & 3.78 (1.5 K)   \\ %\midrule

	\hline\centering

	 32 (0.263) & Fe$_2$PO$_5$ & $Pnma$ (62) & $mmm$ \cite{Fe2PO5s}& I &4.0 \cite{Fe2PO5u} & NO & 3.97, 4.51 & 3.89, 4.22 (12.1 K) \\ %\midrule

	\hline\centering

	33 (0.379) & SmFeO$_3$ & $Pbnm$ (62) & $m^\prime m ^\prime m$ \cite{SmFeO3s}& III&4.0 \cite{TbFeO3u} & YES & 4.48 & 3.692 (300 K) \\ %\midrule

	\hline\centering

	34 (0.336) & NdFeO$_3$ &  $Pnma$ (62) & $m^\prime m ^\prime m$ \cite{NdFeO3s}& III&5.0 \cite{NdFeO3u} & YES& 4.60 & 4.18 (1.5 K)   \\ %\midrule

	\hline\centering

	 35 (0.351) & TbFeO$_3$ & $Pbnm$ (62) & $m^\prime m ^\prime m$ \cite{TbFeO3s}& III&4.0 \cite{TbFeO3u} & YES& 4.46 & 4.8 (4.2 K)   \\ %\midrule

	\hline\centering

	36 (0.380) & SmFeO$_3$ & $Pnma$ (62) & $m^\prime m ^\prime m$ \cite{SmFeO3s}& III&4.0 \cite{TbFeO3u} & YES & 4.58 & 2.836 (500 K)   \\ %\midrule

	\hline\centering
	
	37 (0.402) & Sr$_4$Fe$_4$O$_{11}$ & $Cmmm$ (65) & $m^\prime m ^\prime m$ \cite{Sr4Fe4O11s}& III&4.9 \cite{CoF3u} & YES & 4.67, 0.02 & 3.55, 0 (1.5 K)  \\ %\midrule

	\hline\centering

	38 (0.501) & LiFe$_2$F$_6$ & $P4_2/mnm$ (136) & $4^\prime /mm^\prime m$ \cite{LiFe2F6s}& III&4.0 \cite{LiFe2F6u} & NO& 4.42 & 4.4 (4.2 K) \\ %\midrule

	\hline\centering

	39 (0.116) & FeCO$_3$& $R\bar{3}c$ (167) & $\bar{3}m$ \cite{FeCO3s}& I &5.0 \cite{FeCO3u} & NO & 3.99 &    \\ %\midrule

	\hline\centering
	40 (0.301) & Sr$_2$CoTeO$_6$ & $P2_1/n$ (14) & $2/m$ \cite{Sr2CoTeO6s}& I &4.1 \cite{Sr2CoOsO6u} & YES & 2.97 & 2.10 (4 K)  \\ %\midrule

	\hline\centering

	41 (0.121) & Li$_2$Co(SO$_4$)$_2$ & $P2_1/c$ (14) & $2^\prime/m^\prime$ \cite{Li2CoSO42s}& III&5.7 \cite{Li2CoSO42u} & YES& 3.05 & 3.33 (1.85 K)  \\ %\midrule

	\hline\centering

	42 (0.56) & Ba$_2$CoGe$_2$O$_7$ & $P\bar{4}2_1m$ (113) & $m^\prime m2^\prime$ \cite{Ba2CoGe2O7s}&III &3.0 \cite{Ba3CoIr2O9u} & YES & 2.88 & 2.9 (2.2 K)  \\ %\midrule

	\hline\centering

	43 (0.178) & CoF$_2$&P4$_2$/mnm (136) & $4^\prime/mm^\prime m$ \cite{CoF2s}& III&6.0 \cite{CoF2u} & NO& 3.15 & 2.60 (1.8 K)   \\ %\midrule

	\hline\centering
	
	44 (0.334) & CoF$_3$& $R\bar{3}c$ (167) & $\bar{3}m$ \cite{lee2018weak} &I &5.3 \cite{CoF3u} & NO& 3.92 & 3.21 (RT) \\ %\midrule

	\hline\centering

	45 (0.113) & NiCO$_3$ & $R\bar{3}c$ (167) & $2/m$ \cite{NiCO3s}& I &3.8 \cite{NiCO3u} & YES & 1.83 & 1.94 (4.5 K)  \\ %\midrule

	\hline\centering

	46 (0.45) & La$_2$NiO$_4$ & $P4_2/ncm$ (138) & $m^\prime m ^\prime m$ \cite{La2NiO4s}& III &6.0 \cite{Ba3NiRu2O9u} & YES& 2.03 & 1.68 (4 K)   \\ %\midrule
	\hline\centering
 
	47 (0.36) & NiF$_2$ & $P4_2/mnm$ (136) & $m^\prime m ^\prime m$ \cite{NiF2s}& III &5.0 \cite{NiF2u} & YES & 1.99 &  \\ %\midrule

	\hline\centering

	 48 (0.261) & NiFePO$_5$ & $Pnma$ (62) & $mmm$ \cite{MFePO5s}& I &Ni: 4.6 \cite{PbNiO3u}  & NO& Ni: 1.87  & Ni: 2.03     \\ %\midrule
	&&&&&Fe: 4.9 \cite{Li3Fe2PO43u}&&Fe: 4.58 & Fe: 4.09 (1.5 K)\\

	\hline\centering

	49 (0.21) & PbNiO$_3$ & $R3c$ (161) & $3m$ \cite{PbNiO3u}& I &4.6 \cite{PbNiO3u} & NO& 1.84 &   \\ %\midrule

	\hline\centering

	50 (0.241) & Y$_2$Cu$_2$O$_5$ & Pna2$_1$ (33) & mm2 \cite{Y2Cu2O5s}& I &4.0 \cite{Y2Cu2O5u} & NO & 0.68 & 1.1 (1.5 K)   \\ %\midrule
	\end{tabular}}
	\end{ruledtabular}
\end{table*}
\begin{table*}
\footnotesize
	\caption{\label{tab:table2}Continuation of Table \ref{compounds}}\label{compounds2}
	\begin{ruledtabular}
 \scalebox{0.94}{
	\begin{tabular}{ccccccccc}
 %   \toprule %\midrule
\hline
		S. No & Compound & Structure & MPG &MSG& U value  & CHE   & Local moment & Local moment    \\
	&&&[Reference]&type&(eV)&\footnotesize{symmetry-}&\footnotesize{calculation}&\footnotesize{experiment} \\
	&&&&&& \footnotesize{allowed}&($\mu_B$)&($\mu_B$)\\
	\hline\centering
	51 (0.137) & Cu$_2$V$_2$O$_7$ & $Fdd2$ (43) & $m^\prime m^\prime2$ \cite{Cu2V2O7s}& III &6.52 \cite{Cu2V2O7u} & YES & 0.82 & 0.93 (4 K)  \\ 
 \hline\centering
	52 (0.260) & CuFePO$_5$ & $Pnma$ (62) & $mmm$ \cite{MFePO5s}& I &Cu: 4.0  \cite{Y2Cu2O5u}  & NO& Cu: 0.72  & Cu: 0.95      \\
    &&&&&Fe: 4.9  \cite{CoF3u}&&Fe: 4.58  &Fe: 4.28  \\
    &&&&&&&&(1.4 K)\\
    \hline\centering
	53 (0.239) & Ca$_3$LiRuO$_6$ & $R\bar{3}c$ (167) & $2^\prime/m^\prime$ \cite{Ca3LiRuO6s}& III &2.0 \cite{vsmejkal2020crystal} & YES & 2.05 & 2.8 (4 K)  \\ %\midrule

	\hline\centering

	54 (0.361) & Sr$_3$LiRuO$_6$& $R\bar{3}c$ (167) & $2^\prime/m^\prime$ \cite{Sr3LiRuO6}& III &2.0 \cite{vsmejkal2020crystal} & YES & 2.05 & 2.03 (1.5 K)  \\ %\midrule

	\hline\centering

	55 (0.607) & RuO$_2$$^*$ & $P4_2/mnm$ (136) & $4^\prime/mm^\prime m$ \cite{vsmejkal2020crystal}& III &2.0 \cite{vsmejkal2020crystal} & NO & 0.98 & 0.05 (RT)  \\ %\midrule

	\hline\centering

	56 (0.748) & Ba$_3$NiRu$_2$O$_9$ & $P6_3/mmc$ (194) & $6^\prime/m^\prime mm^\prime $ & III &6.0 (Ni) \cite{Ba3NiRu2O9u}  & NO& Ni:2.02  & Ni:1.7   \\ %\midrule
    &&&\cite{Ba3NiRu2O9s}&&2.0 (Ru) \cite{vsmejkal2020crystal}&&Ru:2.01&Ru:1.5 (5 K)\\
	\hline\centering

	57 (0.434) & K$_2$ReI$_6$ & $P2_1/n$ (14) &  $2/m$ \cite{K2ReI6u}& I &2.0 \cite{K2ReI6u} & YES& 2.67 & 2.2 (1.5 K)  \\ %\midrule

	\hline\centering

	58 (0.210) & Sr$_2$CoOsO$_6$ & $B2/n$ (15) & $2/m$ \cite{Sr2CoOsO6s}& I &4.1 (Co) & YES& Co: 3.28  & Co: 2.6   \\ %\midrule
    &&&& &2.1 (Os) \cite{Sr2CoOsO6u}&&Os: 1.93&Os: 0.7 (2 K)\\
	\hline\centering

	59 (0.3) & Ca$_3$LiOsO$_6$ & $R\bar{3}c$ (167) & $2^\prime/m^\prime$ \cite{Ca3LiOsO6s}& III &2.5 \cite{Ca3LiOsO6u} & YES & 2.28 & 2.0 - 2.3 (4 K)  \\ %\midrule

	\hline\centering

	60 (0.25) & NaOsO$_3$ & $Pnma$ (62) & $m^\prime m^\prime m$ \cite{NaOsO3s}& III &2.0 \cite{NaOsO3u} & YES & 1.97 & 1.0 (200 K)  \\ %\midrule

	\hline\centering

	61 (0.747) & Ba$_3$CoIr$_2$O$_9$ & $C2/c$ (15) & $2/m$ \cite{Ba3CoIr2O9u}& I &3.0 (Co)  & YES& Co:3.08  & Co:2.45  \\ %\midrule
    &&&&&2.4 (Ir) \cite{Ba3CoIr2O9u}&&Ir:1.05&Ir:0.57 (80 K)\\
	\hline\centering

	62 (0.79) & CaIrO$_3$ & $Cmcm$ (63) & $m^\prime m^\prime m$ \cite{CaIrO3s}& III &2.0 \cite{CaIrO3u} & YES& 0.64 &     \\ %\midrule
	      \end{tabular}}
    \end{ruledtabular}
\end{table*}
\vspace{-0.5cm}
\section{Measures for the broken spin degeneracy}\label{sec_measures}
We now introduce quantitative measures to compare the energy splitting between spin-up and spin-down bands.
Since SOC is not included, the Hamiltonian is block diagonal in spin space, so we denote it as $H_{\mathbf{k}\sigma}$.

As a first measure of the spin splitting, we consider the squared difference between bands having opposite spin but the same index. 
We average this quantity over the full Brillouin zone (BZ) and over all occupied states, expressing it as
\begin{equation}
    \Delta_\varepsilon =  \sqrt{\frac{1}{N_{occ} N_k}{\sum_{\mathbf{k},n}(\varepsilon_{n,\mathbf{k},\uparrow}-\varepsilon_{n,\mathbf{k},\downarrow})^2}},
    \label{eq_m1}
\end{equation}
where $N_{occ}$ is the number of occupied  bands and $N_k$ is the number of sampled $k$ points.
We also consider the average energy difference between opposite spin bands at each $k$ point over the entire spectrum (i.e., including unoccupied states). 
Averaging over the BZ as before yields
\begin{equation}
    \Delta_{H} = \sqrt{\frac{1}{N_{b} N_k}{\sum_{\mathbf{k}}\Big(\text{Tr}[H_{\mathbf{k}\uparrow}] -\text{Tr}[H_{\mathbf{k}\downarrow}]\Big)^2}}.
    \label{eq_m2}
\end{equation}
In Eq.~\eqref{eq_m1}, the index $n$ runs over all occupied bands and the eigenvalues of the spin $\uparrow$ and spin $\downarrow$ Hamiltonian are assumed to be ordered. 
Notice that this introduces some arbitrariness with regards to which pairs of opposite-spin bands are taken into account to measure the average energy difference. 
This arbitrariness is absent in Eq.~\eqref{eq_m2}.
The latter, however, is prone to cancellations if pairs of bands are spin split in opposite directions, and suffers from the inaccurate description of the high-lying states. 
The latter is due to the minimal basis set of local orbitals employed in FPLO, which facilitates an accurate description of the occupied states and the low-lying unoccupied states, but gets increasingly inaccurate at higher energies. 
To remedy the inherent drawbacks of Eqs.~\eqref{eq_m1} and \eqref{eq_m2} as measures of spin splitting, we introduce yet another measure, which takes into account the orbital character of the eigenstates:
\begin{equation}
    \Delta_{\varepsilon, \lambda} =  \sqrt{\frac{1}{N_{occ} N_k}{\sum_{\mathbf{k},n}\left(\langle\lambda_{n,\mathbf{k},\uparrow}|\lambda_{n,\mathbf{k},\downarrow}\rangle\left(\varepsilon_{n,\mathbf{k},\uparrow}-\varepsilon_{n,\mathbf{k},\downarrow}\right)\right)^2}},
    \label{eq_m3}
\end{equation}
where $|\lambda_{n,\mathbf{k},\sigma}\rangle$ is the $n^{\rm th}$ eigenvector of spin $\sigma$ and the scalar product only takes into account the orbital sector (i.e. not the spin).
Equation ~\eqref{eq_m3} can be viewed as an extension of Eq.~\eqref{eq_m1}, where the bare difference between the spin-up and spin-down eigenvalues is supplemented with an additional orbital-dependent weight factor. 
This factor is close to unity if the respective eigenvalues in both spin channels have similar orbital characters and tends to zero if their orbital characters are significantly different.
Admittedly, bands of same index and opposite spin could have different energy due to a different orbital character (which would imply, e.g.,  a different crystal field energy). 
The measure $\Delta_{\varepsilon,\lambda}$ aims at discarding the spin-splitting contributions that stem from such kinds of orbital effects. 
Notice that it still suffers from taking energy differences between bands according to their band index. Due to this,  it may be regarded as a lower bound: for a given band, it removes  contributions to the average spin-splitting when neighboring  bands have substantially different orbital character; however, it does not account for the splitting with its partner band (having the most similar orbital character) unless this partner band is indexed immediately below or above it. Last, notice that variations of Eq.~\eqref{eq_m1} taking into account off-diagonal  scalar product ($\langle\lambda_{n,\mathbf{k},\uparrow}|\lambda_{m,\mathbf{k},\downarrow}\rangle$) or, when present, the action of symmetry operations connecting crystallographic sites occupied by opposite-spin ions, could be desirable to fully isolate spin from orbital effects.\\\\
Figure \ref{fig_measures} shows the values of $\Delta_\varepsilon$, $\Delta_H$ and $\Delta_{\varepsilon,\lambda}$ for all the insulating compounds. 
Different color sectors correspond to sets of compounds where the long range magnetic order is associated with different magnetic ions. 
It is apparent that the different measures exhibit certain degree of correlation. 
Indeed, the Pearson correlation coefficient which measures the linear correlation across all compounds amounts to 0.74 between $\Delta_\varepsilon$ and $\Delta_{\varepsilon,\lambda}$ and to 0.5 between $\Delta_{\varepsilon,\lambda}$ and $\Delta_H$.
While no particular magnetic ion can be identified as systematically yielding a particularly sizable spin splitting, several flourides and a sulfate fluoride are among the compounds showing the largest spin splitting (e.g., MnF$_2$,    FeSO$_4$F, LiFe$_2$F$_6$, CoF$_2$ and NiF$_2$). 
This is likely related to the record-high electronegativity of fluorine, which in turn gives rise to largely ionic metal-fluorine chemical bonds. 
A detailed investigation of pertinent magnetostructural correlations is highly desirable, but it is beyond the scope of present work. 
\begin{figure*}
	\begin{center}
		\includegraphics[width=\textwidth]{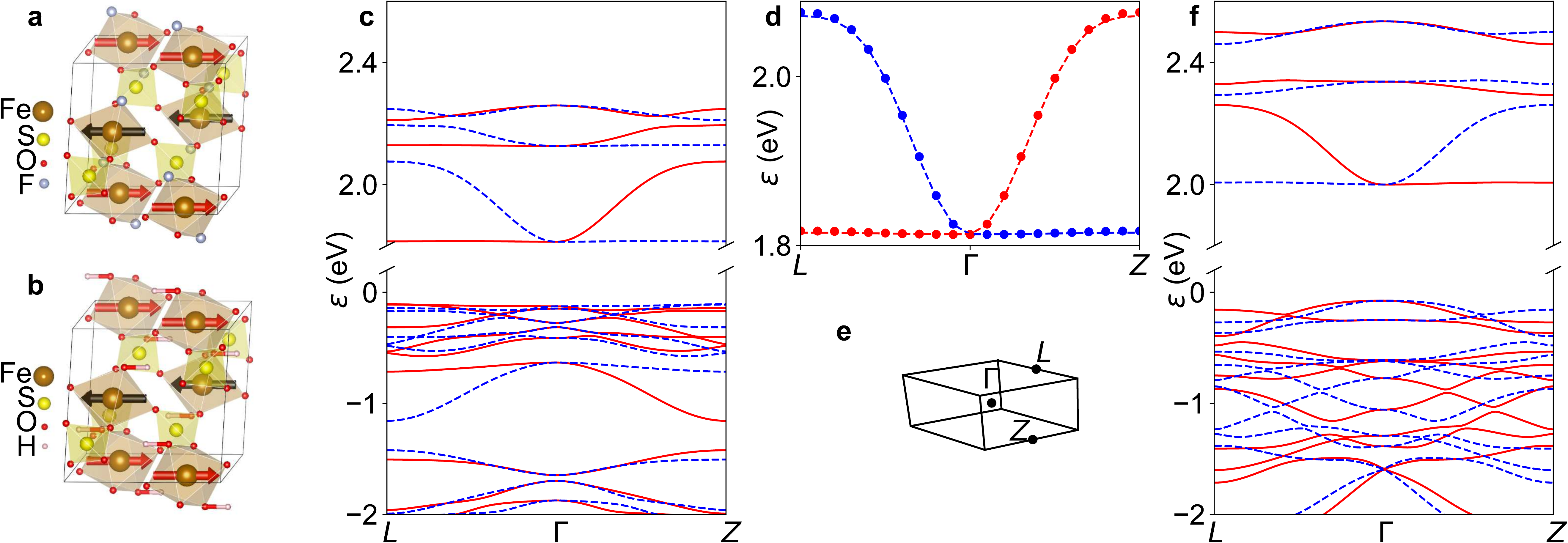}
		\caption{\textbf{FeSO$_4$X family, $X=F$ or $OH$.} (a, b) Crystal structures  of  FeSO$_4$F and FeSO$_4$OH, respectively. (c) Band structure of FeSO$_4$F. The red and blue curves correspond to spin-up and spin-down states, respectively. (d) Fit of the effective model Hamiltonian Eqs. [\ref{eq_efmodel1}-\ref{eq_efmodel3}] to the two lower conducting bands in panel (c). The dots are obtained from DFT, while the dashed lines from the model. (e) Brillouin zone. (f) Band structure of FeSO$_4$OH.} 
		\label{fig_FeSO4}
	\end{center}
\end{figure*}
\vspace{-0.3cm}
\section{Exemplary compounds}
\label{sec_examples}
Here we highlight several compounds which are of interest, either due to their large spin splitting, due to their low-energy electronic structure, or because they are metallic.
\subsection{FeSO$_4$X, X $=\{$F, OH$\}$}\label{FeSO4F}
FeSO$_4$F and FeSO$_4$OH are antiferromagnets with Néel temperatures of 100 K \cite{FeSO4Fs} and 125 K \cite{FeOHSO4s}, respectively.
Both have a centrosymmetric monoclinic crystal structure (space group $C2/c$), with magnetic Fe atoms occupying only one crystallographic site at $\left(\frac14,\frac14,\frac12\right)$. 
The structure of FeSO$_4$F consists of FeO$_4$F$_2$ octahedra connected to each other via F atoms, forming buckled chains [Fig.~\ref{fig_FeSO4}(a)]. 
Oxygen atoms are at the same time bound within SO$_4$ tetrahedra that connect the chains by sharing O atoms. 
Neutron diffraction data showed that atomic magnetic moments align along the $\hat{b}$ axis, the MSG being $C2^\prime/c^\prime$ \cite{FeSO4Fs}. 
The crystal and magnetic structure of FeSO$_4$OH is isostructural to that of FeSO$_4$F [Fig.~\ref{fig_FeSO4}(b)] \cite{FeOHSO4s}. 
Each F atom in the FeO$_4$F$_2$ octahedron is substituted by a hydroxyl group OH$^{-1}$. 

For FeSO$_4$F, the scalar relativistic and nonmagnetic DFT calculation yields a metallic band structure. 
Once the AFM experimental magnetic ordering is considered, a gap of $\sim 1$\,eV opens at the Fermi level. 
Performing GGA$+U$ further increases the band gap to $\sim 1.8$\,eV. 
The calculated magnetic moment of Fe$^{3+}$ is 4.65 $\mu_B$ and matches well the value of 4.32 $\mu_B$ determined from neutron diffraction experiments \cite{FeSO4Fs}. 
FeSO$_4$OH has a similar phenomenology.
Figures \ref{fig_FeSO4}(c) and (d) show the corresponding band structures. 
The energy splitting between opposite-spin states can reach up to $\sim250$\,meV in the lowest conduction band. Some valence bands exhibit an even larger spin-splitting, e.g. of about 400\,meV in FeSO$_4$F for the  pair of bands $\sim1\,$eV below the Fermi energy.

We now consider a simple $2\times2$ Hamiltonian in momentum space to describe the lowest pair of conducting bands. 
In the absence of spin-orbit coupling, the model can be chosen diagonal in a basis of states with well defined spin 
\begin{equation}
H_{\mathbf{k}}=\text{diag}[h_{\mathbf{k}\downarrow},h_{\mathbf{k}\uparrow}].
\label{eq_efmodel1}
\end{equation}
Here, $\mathbf{k}$ is the crystal momentum, which we write in the basis of the primitive vectors $\mathbf{G}_i$ as $\mathbf{k}=2\pi(k_1\mathbf{G}_1+k_2\mathbf{G}_2+k_3\mathbf{G}_3)$.

As explained in Appendix \ref{app_effective_model}, the functions $h_{\mathbf{k}\downarrow}, h_{\mathbf{k}\uparrow}$ can in general be expanded as a Fourier series. 
The expansion coefficients  can be fitted to a pair of bands from DFT, provided these are separated from other bands. 
Here we consider a simple effective model derived in such way. 
The model is defined by Eq.~\eqref{eq_efmodel1} together with
\begin{eqnarray}
h_{\mathbf{k}\uparrow}&=&a_{000}^{\uparrow}+a_{100}^{\uparrow}\cos k_1+a_{010}^{\uparrow}\cos k_2+a_{020}^{\uparrow}\cos 2k_2\nonumber\\
&&+a_{011}^{\uparrow}\cos k_2\cos k_3, \label{eq_efmodel2}\\
h_{\mathbf{k}\downarrow}&=&a_{000}^{\downarrow}+a^{\downarrow}_{010}\cos k_2+a_{100}^{\downarrow}\cos k_1+a_{200}^{\downarrow}\cos 2k_1\nonumber\\
&&+a^{\downarrow}_{101}\cos k_1 \cos k_3.
\label{eq_efmodel3}
\end{eqnarray}
The combination of the time-reversal operator with a twofold rotational symmetry with respect to $\mathbf{q} = \mathbf{G}_1 + \mathbf{G}_2$ or with a reflection with respect to a plane perpendicular to $\mathbf{q}$ are symmetries of both FeSO$_4$F and FeSO$_4$OH. 
These symmetries imply that
\begin{align}
    h_{k_1,k_2,k_3,\sigma}&=h_{-k_2,-k_1,k_3,-\sigma}, \label{eq_sim1}\\
    h_{k_1,k_2,k_3,\sigma}&=h_{k_2,k_1,-k_3,-\sigma}. \label{eq_sim2}
\end{align}
The Hamiltonian Eqs.~\eqref{eq_efmodel1}-\eqref{eq_efmodel3} obeys these symmetries if 
$a_{000}^{\downarrow}=a_{000}^{\uparrow}$, $a_{100}^{\downarrow} = a_{010}^{\uparrow}$,
$a_{010}^{\downarrow}=a_{100}^{\uparrow}$,
$a_{200}^{\downarrow}= a_{020}^{\uparrow}$, and
$a_{101}^{\downarrow}=a_{011}^{\uparrow}$.
Figure \ref{fig_FeSO4}(d) shows the bands resulting from fitting these parameters in the case of FeSO$_4$F. 
The fit was done using data in a mesh over the full Brillouin zone and yields the values (in eV) 
$a_{000}^{\downarrow}\approx0.173$, $a_{100}^{\downarrow}\approx-0.129$, $a_{010}^{\downarrow}\approx-0.005$, $a_{200}^{\downarrow}\approx-0.030$, and
$a_{101}^{\downarrow}\approx 0.004$.

\begin{figure*}
	\begin{center}
		\includegraphics[width=\textwidth]{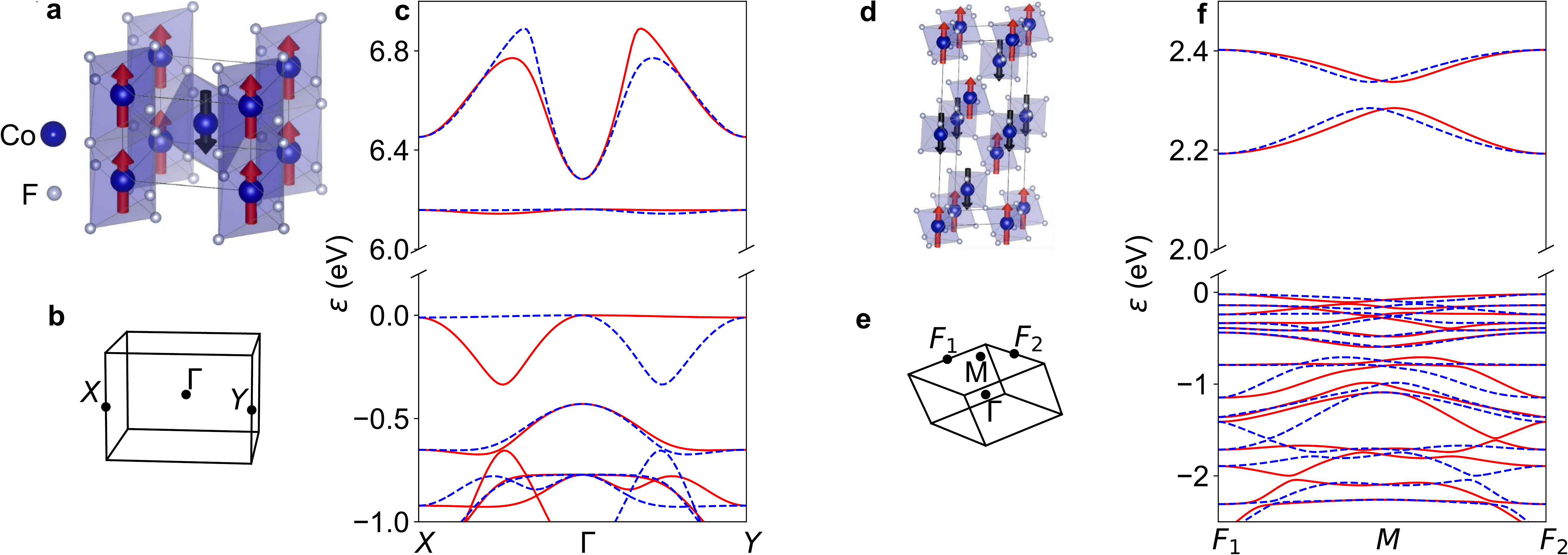}
		\caption{\textbf{CoF$_n$ family, $n=2,3$.} (a, b, c) Crystal structure, Brillouin zone and band structure of CoF$_2$. (d, e, f) Same for CoF$_3$. The red and blue curves correspond to spin-up and spin-down states, respectively. }
		\label{fig_CoF}
	\end{center}
\end{figure*}

\subsection{CoF$_n$, $n=\{ 2,3\}$ }
As shown in Fig.~\ref{fig_measures}, several of the compounds having the largest average spin splitting are transition metal fluorides.
The electronic structure of some of these compounds have been recently investigated with emphasis on the bands' spin splitting (see results on CuF$_2$ in Ref.~\cite{jungwirth2021altermagnetism} and on MnF$_2$ in Refs.~\cite{Yuan2020Giant, Yuan}).
These difluorides crystallize in a tetragonal rutile structure with the space group P4$_2$/mnm while the trifluorides exhibit a $R\bar{3}c$ structure [Fig.~\ref{fig_CoF}(a) and (d)], with the exception of MnF$_3$, which crystalizes in the $C2/c$ space group \cite{hepworth1957crystal, CoF3u}. 

Here we focus on CoF$_2$ and CoF$_3$. The former orders antiferromagnetically at 38 K \cite{CoF2s} and the latter at 460 K \cite{Wollan1958, lee2018weak}.
The magnetic moments align with the polar axis such that the MSG is $4^\prime/mm^\prime m$ and $3m$, respectively.
This magnetic anisotropy is different from that of other fluorides, such as NiF$_2$ and FeF$_3$, where the magnetic moments point in plane, breaking some of the rotation symmetries. 

Figures \ref{fig_CoF}(c) and (f) show the band structures of CoF$_2$ and of CoF$_3$, respectively. 
The difluoride compound has a rather clean band structure below the Fermi level and the highest occupied bands reach a spin splitting as large as 300\,meV.

\begin{figure}
  \centering
	\begin{center}
		\includegraphics[width=1.0\columnwidth]{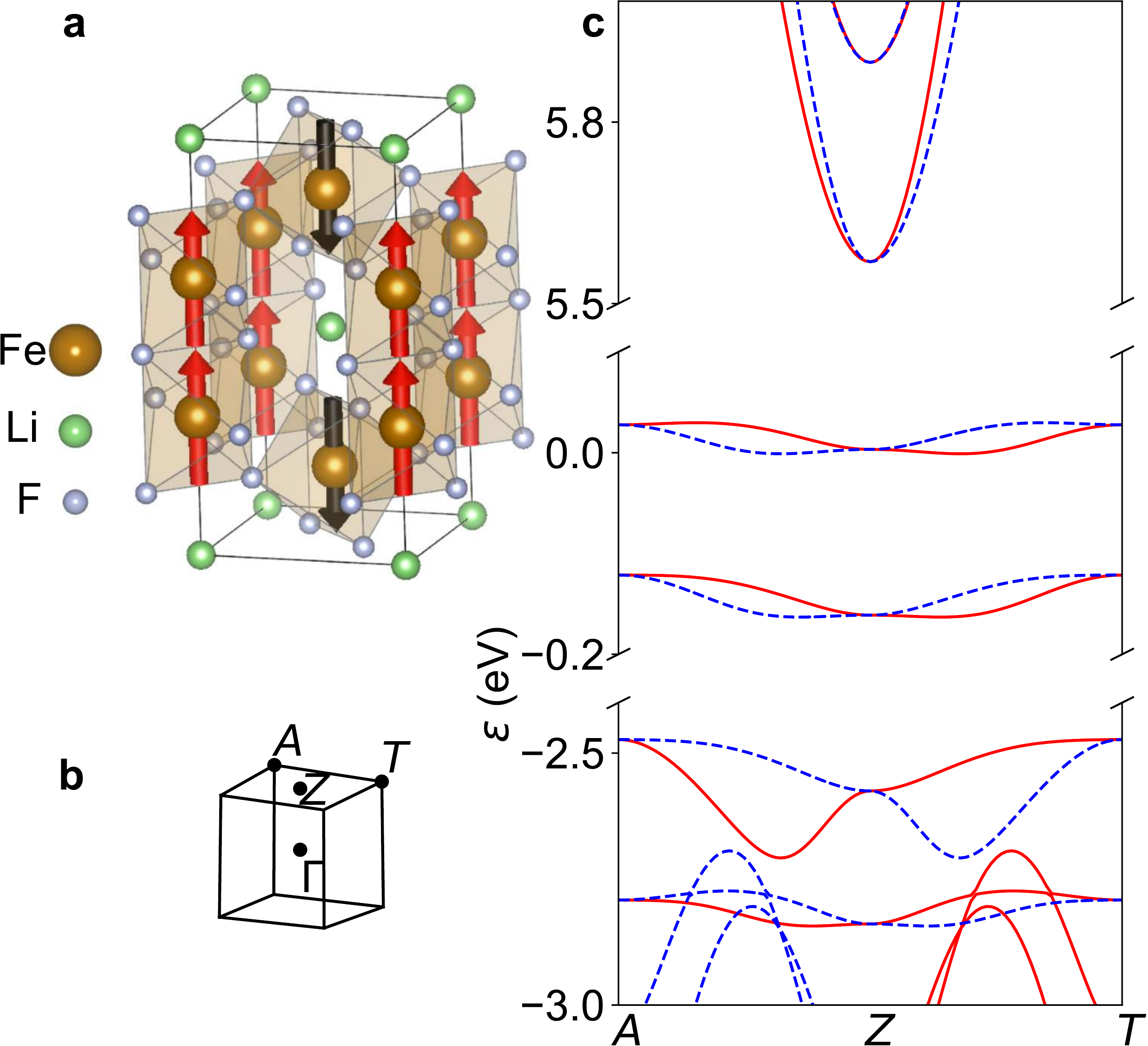}
		\caption{\textbf{Spin-split antiferromagnetic flat bands. } (a, b, c) Crystal structure, Brillouin zone and band structure of LiFe$_2$F$_6$. The red and blue curves correspond to spin-up and spin-down states, respectively. }
		\label{fig_LiFe2F6}
	\end{center}
\end{figure}

\subsection{LiFe$_2$F$_6$}
LiFe$_2$F$_6$ has a tetragonal symmetry and two crystal structures have been refined: Ref.~\cite{LiFe2F6s} reported an inversion symmetric space group $P4_2/mnm$ and subsequent work has identified a structure of lower symmetry (space group $P4_2nm$) \cite{fourquet1988trirutile}. Recent theoretical work has suggested that LiFe$_2$F$_6$ is a multiferroic compound \cite{LiFe2F6u}. The experimental direction of magnetic moments in LiFe$_2$F$_6$ is along the crystallographic c-axis and, using the inversion-symmetric description, the corresponding magnetic space group is $P4_2^\prime/mnm^\prime$ \cite{LiFe2F6s}.  
Figure \ref{fig_LiFe2F6}(c) shows the band structure for the centrosymmetric crystal and magnetic structure  
reported in Ref.~\cite{LiFe2F6s}. 
Two pairs of bands are positioned near the Fermi level, separated by $\sim 2$\,eV from other valence states and by $\sim 5$\,eV from the next unoccupied levels. 
Their bandwidth is $\sim 30$\,meV, which is of the same order as their maximum spin splitting. 
The rather flat and spin-split bands which are well separated from other conducting or valence states may be of interest for the ongoing study of materials with flat bands near the Fermi energy \cite{kang2020dirac, yin2020quantum}.

\subsection{Metallic systems: RuO$_2$, CrNb$_4$S$_8$and CrSb }
According to our DFT calculations, among the more than sixty compounds in Tables \ref{compounds} and  \ref{compounds2} only CrNb$_4$S$_8$, RuO$_2$, and CrSb are metallic. 
We note in passing the special case of Sr$_2$CoOsO$_6$, which converges to metallic state without SOC but it becomes insulating state in the fully-relativistic case \cite{Sr2CoOsO6u}. 

The special subset of metallic compounds is of potential interest as a platform for the study of various low-energy transport properties such as the anomalous Hall effect (AHE), which accounts for a finite Hall conductivity at zero external magnetic field. 
In the case of cAFMs, the AHE has been named \textit{crystal Hall effect} (CHE) since it arises from the asymmetric crystal environment of the magnetic atoms having opposite magnetic moment \cite{vsmejkal2020crystal}. 
This asymmetry is reflected in the breaking of both $\theta$ as well as of symmetries involving combinations of $\theta$ with operations that reverse the magnetic order by connecting the two opposite-spin sublattices. 

cAFMs with such kind of combined symmetries, e.g. CuMnAs (which has $\theta I$) or GdPtBi (which has $\theta T$, with $T$ a translation symmetry), have vanishing anomalous Hall effect in linear response.
From this point of view, the CHE and the spin splitting without SOC are closely related. The CHE has, however, additional requirements.
First, the intrinsic AHE arising from the Berry curvature of Bloch states has two contributions: a Fermi-sea part that is nonzero only for topological systems whose occupied band structure carries a net Chern number, and a Fermi surface part \cite{Haldane2004}.
Thus, putting aside topological systems, a Fermi surface is another requirement for the CHE.
Second, some MSGs that allow for spin-split bands still yield a vanishing CHE due to the presence of certain point group symmetries \cite{vsmejkal2020crystal}.

\begin{figure*}
	\begin{center}
		\includegraphics[width=\textwidth]{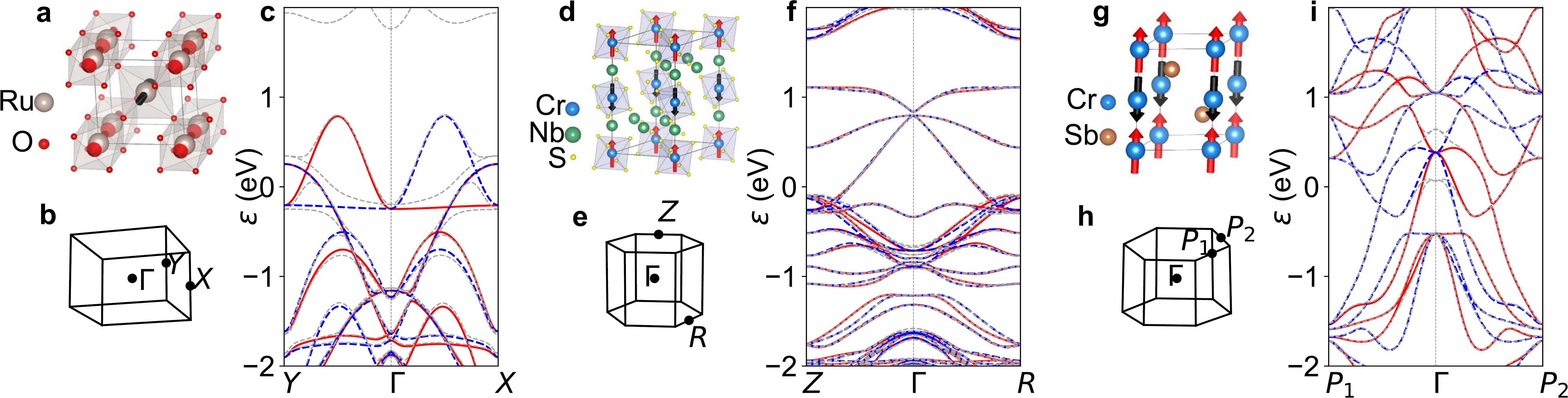}
		\caption{\textbf{Spin-split antiferromagnetic metals.} (a, b, c) Crystal structure, Brillouin zone and band structure of RuO$_2$. The red and blue curves correspond to spin-up and spin-down states, respectively. Dashed gray curves are the bands computed including the spin-orbit coupling. (d, e, f) Same for CrNb$_4$S$_8$. (g, h, i) Same for CrSb.}
		\label{fig_metals}
	\end{center}
\end{figure*}

One compound that was first identified as a possible candidate to study the CHE is RuO$_2$ \cite{vsmejkal2020crystal} and first experimental results have been reported \cite{feng2020observation}. 
The computed electronic structure, shown in Fig.~\ref{fig_metals}(c) and further discussed in the next section, presents a large spin splitting, i.e. $\sim 1$\,eV, particularly around the Fermi level. We find that its average spin splitting as measured by Eq.~\eqref{eq_m2} is the largest among the metallic compounds, followed by CrSb.

Another metallic compound in our material list is CrNb$_4$S$_8$, which belongs to the family of compounds often named $A_xB$S$_2$ where $A$ is a magnetic transition metal such as Mn or Cr, $B$ is either Nb or V and $x=\{1/3,1/4\}$. These compounds consist of layers of NbS$_2$ intercalated by a magnetic transition metal atoms [Fig.~\ref{fig_metals}(d)].
In Ref.~\cite{CrNb4S8s}, for $A=$ Cr, samples with the composition $x\sim1/4$ were reported and their magnetic order described as being composed of ferromagnetic Cr layers coupled to each other antiferromagnetically. The case $x\sim1/3$ was initially described as ferromagnetic \cite{hulliger1970magnetic}, and later argued to present a helical structure \cite{moriya1982evidence} in which the magnetization of neighboring ferromagnetic layers exhibits a slow relative rotation. 
This structure has been further analyzed in subsequent works \cite{Togawa2012, Ghirmire2013} and even the presence of magnetic field-induced solitons has been indicated \cite{Togawa2012,osorio2021creation}. 
We are unaware of similar investigations for samples with the lower nominal composition $x=1/4$.

Figure \ref{fig_metals}(f) shows the computed band structure based on the AFM structure reported in Ref.~\cite{CrNb4S8s}. 
The spin splitting is modest as compared to RuO$_2$. 
Notice, in addition, that for this compound the CHE vanishes due to the threefold rotation symmetry. 
The related compound CoNb$_3$S$_6$ has been predicted to host a finite CHE in Ref.~\cite{CoNb3S6}. 
The difference stems from the easy-plane magnetic anisotropy in the latter compound, which lowers the symmetry.
 
Lastly, CrSb crystallizes in the hexagonal NiAs structure (SG No.~194), it orders magnetically below 700 K \cite{Snow1952, Takei1963}, and it is also a metallic compound. It has been found of interest for different reasons, including the possibility of using it in heterostructures of topological insulators \cite{he2017tailoring, He2018}. 
Its electronic structure has been studied in several works \cite{dijkstra1989electronic, ito2007electronic, polesya2011structural, Park2020}.
The structure is centrosymmetric and consists of a stack of triangular lattices of Cr and of Sb. The Cr ions order ferromagnetically within planes and antiferromagnetically between planes [Fig.~\ref{fig_metals}(g)]. The spin-splitting can be traced to the relative stacking of the triangular lattices: for each Cr plane,  the Sb triangular lattices that are placed above and below it are rotated by an angle $2\pi/6$ with respect to each other. As a result, opposite-spin Cr ions are related by the symmetry $C_6 \Theta+(0,0,c/2)$  which, together with inversion symmetry, does not enforce spin-degeneracy in a given $k$-point but at $k$-points related by $2\pi/6$. Notice that the essentially non-magnetic Sb planes play a central role in the momentum-space spin-splitting.
We find that the average spin splitting as measured by Eq.~\eqref{eq_m2} is the second largest among the metallic compounds. 
As in CrNb$_4$S$_8$, the AHE vanishes but it would become finite with an easy-plane magnetic anisotropy, something that has been recently predicted for this compound under uniaxial strain \cite{Park2020}.

Overall, although a rather large number of compounds in the MAGNDATA database allow for the CHE from a symmetry point of view (as indicated in Tables \ref{compounds} and \ref{compounds2}), the requirement of a Fermi surface makes the available compounds showing this effect a rather special subset. 
We have found only three metallic compounds, all of them having  vanishing CHE in their magnetic ground state 
due to either the fourfold or the sixfold rotational symmetry. Notice that as already experimentally shown in RuO$_2$, a field-induced reorientation of the N\'{e}el vector can readily enable the observation of the CHE \cite{feng2020observation}.
One important class that we have filtered out from the MAGNDATA database are rare-earth-based magnetic compounds. 
Often, in such systems other electronic shells are responsible for the compound being metallic and it could be the case that the combination of a Fermi surface together with the required symmetry conditions for the CHE is more abundant, as already predicted in some rare-earth-based Weyl semimetals \cite{ray2022tunable}. Last, we note in passing that other metallic collinear antiferromagnets have been recently discussed, such as Mn$_5$Si$_3$ \cite{Libor2022PRX} and FeSb$_2$ \cite{mazin2021prediction}.

\section{SPIN-ORBIT COUPLING EFFECTS}

The highlight of spin-split antiferromagnets is the possibility of having spin-momentum textures in the absence of SOC. On the other hand, we have built our materials database based on the classification provided by magnetic space groups, which are founded on symmetry operations that by construction assume the SOC  to be relevant. Indeed, the SOC generally contributes to the magnetic anisotropy which in turn affects the magnetic space group associated to a given system. We expect, however, that whether or not the bands of a system present spin-splitting in the limit of zero SOC does not depend on the direction of the magnetic moments in real space. In fact, in that limit the scalar-relativistic calculations presented above are independent of this direction.

Naturally, the SOC does affect the spin-momentum textures, depending on its strength. In systems where all elements are relatively light, like FeSO$_4$F, the effects of the SOC are strongly confined near band crossings that become avoided in the fully-relativistic calculation. In systems with heavy elements, like RuO$_2$ or CrSb, the effect of SOC is less obvious. To illustrate these aspects, we now describe the case of RuO$_2$.
\begin{figure*}
  \centering
	\begin{center}
		\includegraphics[width=1.0\textwidth]{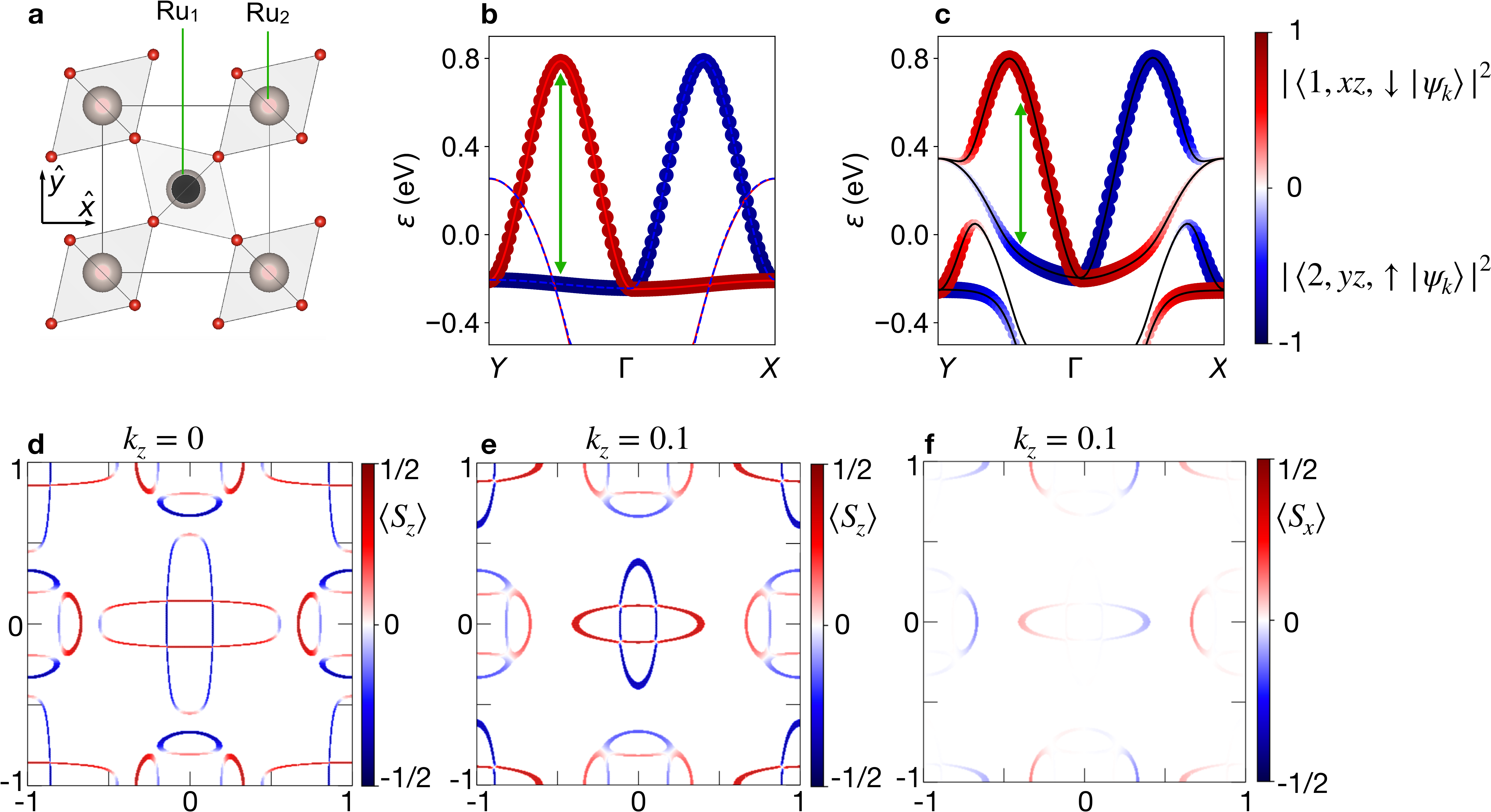}
		\caption{\textbf{Spin-orbit coupling effects in RuO$_2$. } (a) RuO$_2$ crystal structure. (b,c) Band structure obtained from  scalar-relativistic and fully-relativistic calculations, respectively. The dots size and color indicates the projection of the corresponding Bloch eigenstates on the 4d $xz$ orbital spin $\downarrow$ of Ru at site 1 or the 4d $yz$ spin $\uparrow$ orbital of Ru at site 2. (d,e,f) Spin texture at cuts of the Fermi surface, (d) shows the expectation value of $S_z$ at $k_z=0$, (e) and (f) those of $S_z$ and $S_x$ at $k_z=0.1$. Momenta are presented in units of $2\pi/a$.}
		\label{fig_RuO2soc}
	\end{center}
\end{figure*}
In RuO$_2$, the two opposite-spin Ru ions per unit cell occupy crystallographic sites related by the screw $C_4+(\frac{1}{2},\frac{1}{2},\frac{1}{2})$, as illustrated in Fig. \ref{fig_RuO2soc}a.
Figure \ref{fig_RuO2soc}b shows the scalar relativistic band structure spin-resolved and projected on relevant 4d-orbitals of Ru atoms as explained below.
Four bands cross the Fermi energy, a pair of spin-degenerated bands and a pair with a large spin-splitting, which reaches a maximum of 1eV. The latter pair corresponds to bands of predominantly 4d-$xz$ of Ru at site 1 (spin down) and of 4d-$yz$ of Ru at site 2 (spin up). 
 The color and size of the dots in the figure indicate the projection of the corresponding Bloch state on these atomic-like orbitals.
 Notice that since the the oxygen octahedra around the two Ru sites are related by a $C_4$ rotation with respect to the $\hat{z}$ axis, the chosen orbitals at the two Ru sites have the same local orientation with respect to the corresponding local oxygen environment. Hence, energy splittings between Bloch states dominated by these local orbitals do not originate from crystal field effects but from their different spins.
  
Figure \ref{fig_RuO2soc}c shows the fully-relativistic band structure as obtained with the magnetic moments along $\hat{z}$. 
 The main effect of the SOC is to produce an avoided crossing of the spin-degenerated bands with the lower-in-energy, spin-split band. 
As a result, the double degeneracy of the former is lost.
In addition,  the spin-split band involved in the avoided crossing acquires a stronger spin-orbital entanglement close to the avoided crossing.  
As a result, the pure-spin splitting remains an accurate description close to $\Gamma$, while the separation between spin and orbital effects is naturally less defined as the avoided crossing is approached. Still, as illustrated by the green arrow in \ref{fig_RuO2soc}c, rather close to the avoided crossing, a band splitting of around 0.7\,eV between states having $\langle S_z \rangle \simeq \pm \hbar/2$ and similar orbital character can be observed.

It is instructive to analyze the spin-momentum texture at the Fermi surface. Figure \ref{fig_RuO2soc}d shows the Fermi surface contour corresponding to $k_z=0$ in the fully-relativistic approximation. As expected from the discussion above, the spin character of the states in the inner electron pocket is largely unaffected by the SOC, $\langle S_z \rangle \simeq \pm \hbar/2$, in agreement with the results in Ref.~\cite{jungwirth2021altermagnetism}. On the other hand, Bloch states on the outer electron pocket contours present a change in the sign of $\langle S_z \rangle$ as the avoided crossing is approached. Some states thus show $\langle S_z \rangle \simeq 0$, pointing to a strong coupling with orbital degrees of freedom. This is also the case for states near avoided crossings between the hole pockets around the points $X$ and $Y$.
At $k_z \neq 0$ a similar phenomenology is observed (Fig. \ref{fig_RuO2soc}e).
Notice that for states with $k_z=0$, other components of the spin have null expectation value due to the reflection symmetry $M_z: z \to -z$. At finite $k_z$, this is no longer the case and the spin-momentum textures also acquire an in-plane spin component as illustrated in Fig. \ref{fig_RuO2soc}f. The expectation values attained by these components of the spin are, nevertheless, rather small when the magnetic moments point along the easy-axis of the compound.
\section{CONCLUSIONS}
We have performed band structure calculations of more than sixty collinear antiferromagnets which exhibit spin-split electronic bands in the absence of spin-orbit coupling. From the perspective of spin-space groups, this kind of compounds have recently been dubbed altermagnets, a name that reflects the alternating spin-splitting they present in momentum space.
In order to compare the strength of this effect among all the considered materials, we have introduced and discussed different quantitative measures for the average spin splitting. 
This has allowed to identify two families of compounds that systematically show the largest splitting between opposite-spin bands: difluorides MF$_2$ with, e.g., M$=$Co or Ni and FeSO$_4$X, with X$=$F or OH. 

For this latter class, we have introduced a minimal $2\times2$ tight-binding model that describes its two lowest conduction bands. This modeling approach has the advantage of being completely general: in the absence of spin-orbit coupling, two-orbital tight-binding models can be introduced in the way we have described for any pair of spin-split bands of any material, provided that they are sufficiently well-separated from other band pairs.
Given their small size, especially relative to the multi-band Wannier Hamiltonian, these effective Hamiltonians may prove useful when analyzing the behavior of collinear antiferromagnets in real space.
Examples include studying the effect of disorder, or predicting spin-resolved transport properties in realistic, multi-terminal geometries.
We plan to tackle both of these research directions in the future.

Our calculations also led us to highlight some compounds which are of interest due to their peculiar low-energy electronic structure. Here LiFe$_2$F$_6$ stands out for showing low-energy flat spin-split bands and RuO$_2$, CrNb$_4$S$_8$, and CrSb for being metallic. 
We hope our results will contribute to the identification of suitable compounds that facilitate  experimental insights into the physics of spin-split electronic bands in collinear antiferromagnets.%in the limit of vanishing spin-orbit coupling.

\appendix

\section{APPENDIX: 2 $\times$ 2 effective model.}
\label{app_effective_model}
In this Appendix, we present a method to derive two-band models in the limit of vanishing spin-orbit coupling. 
In this limit, the Hamiltonian is diagonal in a basis of states with well-defined spin. 
The procedure consists in Fourier expanding the energy bands $\varepsilon_{\sigma}(\mathbf{k})$ and in fitting the DFT data to estimate the expansion coefficients.

The Fourier series for a one-variable periodic function $O(x)$ reads
\begin{eqnarray}
O(x)=\sum_{n=0}^{\infty}[a_n\cos\frac{2\pi nx}{T} + b_n \sin\frac{2\pi nx}{T}].\label{eq:1dfourier}
\end{eqnarray}
The $a_0$ term corresponds to an overall constant, whereas all other coefficients are obtained by
\begin{eqnarray}
a_n&=&\frac{2}{T}\int_{T}O(x)\cos\frac{2\pi nx}{T}dx,\\
b_n&=&\frac{2}{T}\int_{T}O(x)\sin\frac{2\pi nx}{T}dx.
\end{eqnarray}
Here, we have used the property
\begin{eqnarray}
\frac{1}{T}\int_{T}e^{i2\pi(n-m)x/T}dx=\delta_{nm},
\end{eqnarray}
which leads to the orthogonality of the trigonometric functions, for instance,
\begin{eqnarray}
\int_{T}\cos\frac{2\pi nx}{T}\sin\frac{2\pi mx}{T}=0.
\end{eqnarray}

The bands $\varepsilon_{\sigma}(\mathbf{k})$ are functions of three variables, here taken to be the projections $k_i$ of a given momentum $\mathbf{k}$ on the three primitive vectors $\mathbf{G}_i$. 
Analogously to the one-variable case, they can be expanded in a orthogonal basis as
\begin{eqnarray}
\varepsilon_{\uparrow,\downarrow}(\mathbf{k})&=&\sum_{m,n,v=0}^{\infty}[a_{mnv}\cos 2\pi mk_1\cos 2\pi nk_2\cos 2\pi vk_3]\nonumber\\
&&+b_{mnv}\sin 2\pi mk_1\cos 2\pi nk_2\cos 2\pi vk_3\nonumber\\
&&+c_{mnv}\cos 2\pi mk_1\sin 2\pi nk_2\cos 2\pi vk_3\nonumber\\
&&+d_{mnv}\sin 2\pi mk_1\sin 2\pi nk_2\cos 2\pi vk_3\nonumber\\
&&+e_{mnv}\cos 2\pi mk_1\cos 2\pi nk_2\sin 2\pi vk_3\nonumber\\
&&+f_{mnv}\sin 2\pi mk_1\cos 2\pi nk_2\sin 2\pi vk_3\nonumber\\
&&+g_{mnv}\cos 2\pi mk_1\sin 2\pi nk_2\sin 2\pi vk_3\nonumber\\
&&+h_{mnv}\sin 2\pi mk_1\sin 2\pi nk_2\sin 2\pi vk_3],\nonumber\\
\label{eq:fit_function}\end{eqnarray}
where, $k_i\in[-0.5, 0.5)$.

Provided that the pair of bands to be modeled are separated from other bands, one can obtain the coefficients of the Fourier expansion via a fit to the DFT data corresponding to such a pair of bands. 
This is found to be approximately the case for the lowest conducting bands in FeSO$_4$F.
In principle, Eq.~\eqref{eq:fit_function} can provide a precise model if enough parameters are included. 
We have first kept coefficients up to the fourth-nearest hopping order ($m+n+v\leq 4$). 
We found that the components which are odd in momentum are zero within numerical precision ($\approx 10^{-15}$). 
Thus, we simplify the fit, setting $b_{mnv}=c_{mnv}=e_{mnv}=h_{mnv}=0$. 
The goodness of the fit can be estimated as
\begin{equation}
\xi=\frac{1}{2W}\sqrt{\frac{1}{N}\sum_{\mathbf{k},\sigma}\Big(\varepsilon_{\sigma}(\mathbf{k})-\varepsilon^{\text{DFT}}_{\sigma}(\mathbf{k})\Big)^2.} 
\end{equation}
where $\varepsilon^{\text{DFT}}_{\sigma}$ is the data obtained from DFT and $W\approx 0.27$\,eV is the bandwidth [see Fig.~\ref{fig_FeSO4}(d)].
For FeSO$_4$F, keeping up to fourth-nearest hoppings (which yields 65 parameters) and fitting $N\approx 9000$ points in the full Brillouin zone, we obtain $\xi\approx 0.0007$. 

The resulting fit is only composed of even functions of momentum and, in addition, satisfies 
\begin{align}
    a_{mnv}^{\downarrow}&=a_{nmv}^{\uparrow},\\
    d_{mnv}^{\downarrow}&=d_{nmv}^{\uparrow},\\
    g_{mnv}^{\downarrow}&=-f_{nmv}^{\uparrow},\\
    f_{nmv}^{\downarrow}&=-g_{mnv}^{\uparrow},
\end{align}
within error bars.
Combined with Eq.~\eqref{eq:fit_function}, this reflects the fact that FeSO$_4$F obey inversion symmetry and Eqs.~\eqref{eq_sim1} and \eqref{eq_sim2}, which correspond to the symmetries $C_{110}'$ and $\sigma_{db}'$, respectively.

Last, for the effective model Eqs.~\eqref{eq_efmodel2} and \eqref{eq_efmodel3}, which has only five independent parameters, the goodness of the fit is $\sim 0.008$.

\section*{DATA AVAILABILITY}
All results generated and/or analysed in this study are included in this paper and additional information is  provided in the data repository\\ \href{https://doi.org/10.5281/zenodo.6675680}{https://doi.org/10.5281/zenodo.6675680}.

\section*{CODE AVAILABILITY}
All scripts for calculating the presented results can be provided by the corresponding author upon reasonable request.

\section{ACKNOWLEDGEMENTS}
We thank Ulrike Nitzsche for technical assistance. Y.G. thanks Prof. Karsten Albe from Technical University Darmstadt for supervising the master thesis. We  acknowledge financial support from the German Research Foundation (Deutsche Forschungsgemeinschaft, DFG) via SFB1143 Project No. A5 and under Germany’s Excellence Strategy through Würzburg-Dresden Cluster of Excellence on Complexity and Topology in Quantum Matter—ct.qmat (EXC 2147, Project No. 390858490). J.I.F. would like to thank the support from the Alexander von Humboldt Foundation during the part of his contribution to this work done in Germany, Pablo S. Cornaglia for useful discussions and ANPCyT grants PICT 2018/01509 and PICT 2019/00371.
\section{AUTHOR CONTRIBUTIONS}
Y.G. performed all the \textit{ab-initio} calculations. 
H.L. and C.F. constructed the tight-binding models. 
Y.G. and O.J. computed the different measures to estimate the average momentum-space spin-splitting. 
J.I.F. and O.J. formulated the original idea and supervised  the project together with J.v.d.B. 
 Y.G. and J.I.F. wrote a first version of the manuscript and all authors actively contributed to the analysis of the results as well as to reading and editing the manuscript.
\section{COMPETING INTERESTS}
The authors declare no competing interests.
\renewcommand{\refname}{ REFERENCES}
%\nocite{*}
\bibliographystyle{apsrev4-2}
%\printbibliography
\bibliography{apssamp}% Produces the bibliography via BibTeX.
\end{document}